\documentstyle[12pt,epsfig]{article}

\catcode`\@=11
\long\def\@makefntext#1{
\protect\noindent \hbox to 3.2pt {\hskip-.9pt  
$^{{\ninerm\@thefnmark}}$\hfil}#1\hfill}                

\def\@makefnmark{\hbox to 0pt{$^{\@thefnmark}$\hss}}  
        
\def\ps@myheadings{\let\@mkboth\@gobbletwo
\def\@oddhead{\hbox{}
\rightmark\hfil\ninerm\thepage}   
\def\@oddfoot{}\def\@evenhead{\ninerm\thepage\hfil
\leftmark\hbox{}}\def\@evenfoot{}
\def\sectionmark##1{}\def\subsectionmark##1{}}

\newcounter{sectionc}\newcounter{subsectionc}\newcounter{subsubsectionc}
\renewcommand{\section}[1] {\vspace*{0.6cm}\addtocounter{sectionc}{1} 
\setcounter{subsectionc}{0}\setcounter{subsubsectionc}{0}\noindent 
        {\normalsize\bf\thesectionc. #1}\par\vspace*{0.4cm}}
\renewcommand{\subsection}[1] {\vspace*{0.6cm}\addtocounter{subsectionc}{1} 
        \setcounter{subsubsectionc}{0}\noindent 
        {\normalsize\it\thesectionc.\thesubsectionc. #1}\par\vspace*{0.4cm}}
\renewcommand{\subsubsection}[1]
{\vspace*{0.6cm}\addtocounter{subsubsectionc}{1}
        \noindent {\normalsize\rm\thesectionc.\thesubsectionc.\thesubsubsectionc. 
        #1}\par\vspace*{0.4cm}}

\newcounter{appendixc}
\newcounter{subappendixc}[appendixc]
\newcounter{subsubappendixc}[subappendixc]

\renewcommand{\appendix}[1] {\vspace*{0.6cm}
        \refstepcounter{appendixc}
        \setcounter{figure}{0}
        \setcounter{table}{0}
        \setcounter{equation}{0}
        \renewcommand{\thefigure}{\Alph{appendixc}.\arabic{figure}}
        \renewcommand{\thetable}{\Alph{appendixc}.\arabic{table}}
        \renewcommand{\theappendixc}{\Alph{appendixc}}
        \renewcommand{\theequation}{\Alph{appendixc}.\arabic{equation}}
        \noindent{\bf Appendix \theappendixc #1}\par\vspace*{0.4cm}}

\def\abstracts#1{{
        \centering{\begin{minipage}{12.2truecm}\footnotesize\baselineskip=12pt\noindent
        \centerline{\footnotesize ABSTRACT}\vspace*{0.3cm}
        \parindent=0pt #1
        \end{minipage}}\par}} 

\renewenvironment{thebibliography}[1]
        {\begin{list}{\arabic{enumi}.}
        {\usecounter{enumi}\setlength{\parsep}{0pt}
\setlength{\leftmargin 1.25cm}{\rightmargin 0pt}
         \setlength{\itemsep}{0pt} \settowidth
        {\labelwidth}{#1.}\sloppy}}{\end{list}}

\newcounter{itemlistc}
\newcounter{romanlistc}
\newcounter{alphlistc}
\newcounter{arabiclistc}

\newcommand{\fcaption}[1]{
        \refstepcounter{figure}
        \setbox\@tempboxa = \hbox{\footnotesize Fig.~\thefigure. #1}
        \ifdim \wd\@tempboxa > 6in
           {\begin{center}
        \parbox{6in}{\footnotesize\baselineskip=12pt Fig.~\thefigure. #1}
            \end{center}}
        \else
             {\begin{center}
             {\footnotesize Fig.~\thefigure. #1}
              \end{center}}
        \fi}

\newcommand{\tcaption}[1]{
        \refstepcounter{table}
        \setbox\@tempboxa = \hbox{\footnotesize Table~\thetable. #1}
        \ifdim \wd\@tempboxa > 6in
           {\begin{center}
        \parbox{6in}{\footnotesize\baselineskip=12pt Table~\thetable. #1}
            \end{center}}
        \else
             {\begin{center}
             {\footnotesize Table~\thetable. #1}
              \end{center}}
        \fi}

\def\@citex[#1]#2{\if@filesw\immediate\write\@auxout
        {\string\citation{#2}}\fi
\def\@citea{}\@cite{\@for\@citeb:=#2\do
        {\@citea\def\@citea{,}\@ifundefined
        {b@\@citeb}{{\bf ?}\@warning
        {Citation `\@citeb' on page \thepage \space undefined}}
        {\csname b@\@citeb\endcsname}}}{#1}}

\newif\if@cghi
\def\cite{\@cghitrue\@ifnextchar [{\@tempswatrue
        \@citex}{\@tempswafalse\@citex[]}}
\def\citelow{\@cghifalse\@ifnextchar [{\@tempswatrue
        \@citex}{\@tempswafalse\@citex[]}}
\def\@cite#1#2{{$\null^{#1}$\if@tempswa\typeout
        {IJCGA warning: optional citation argument 
        ignored: `#2'} \fi}}

 1
 1
 1

\font\ninerm=cmr9

\newcommand{\etal}{{\em et al\/}\ }
\newcommand{\newref}{\\\hspace*{-12pt}}

\begin{document}
\begin{flushright}
  BI-TP 97/50\\ CERN-TH/97-322\\ WUE-ITP-97-044\\[1ex] 
  {\tt hep-ph/9711352}\\ 
\end{flushright}
\vskip 35pt

\centerline{\normalsize\bf LEPTOQUARKS AND THE HERA HIGH-$Q^2$   
  EVENTS\footnote{Talk presented by R.\ R\"uckl at the Ringberg
    Workshop on Physics beyond the Standard Model, {\it Beyond the
      Desert}, June 8 -- 14, 1997, Tegernsee, Germany.} } 
\baselineskip=22pt

\centerline{\footnotesize R.\ R\"UCKL}
\baselineskip=13pt
\centerline{\footnotesize\it Institut f\"ur Theoretische Physik,
  Universit\"at W\"urzburg}
\baselineskip=12pt
\centerline{\footnotesize\it D-97074 W\"urzburg, Germany}
\baselineskip=12pt
\centerline{\footnotesize\it and}
\baselineskip=12pt
\centerline{\footnotesize\it Theory Division, CERN, CH-1211, 
  Gen\`{e}ve 23, Switzerland} 
\vspace*{0.3cm}
\centerline{\footnotesize and}
\vspace*{0.3cm}
\centerline{\footnotesize H.\ SPIESBERGER}
\baselineskip=13pt
\centerline{\footnotesize\it Fakult\"at f\"ur Physik, Universit\"at
  Bielefeld, D-33501 Bielefeld, Germany}
\vspace*{0.9cm} 
\abstracts{
The excess of high-$Q^2$ events recently observed in deep-inelastic
positron-proton scattering at HERA has refuelled speculations on physics
beyond the standard model, in particular on low-mass leptoquark-type
particles. We review the theoretical framework for leptoquark
interactions, and their production and decay at HERA.  Bounds on
leptoquark masses and couplings, and implications on other experiments
are also discussed.}

\normalsize\baselineskip=15pt

\section{The data}

Both HERA experiments, H1 \cite{H1data} and ZEUS \cite{Zdata}, have
reported the observation of an excess of events in deep-inelastic
positron-proton scattering at large values of Bjorken-$x$ and momentum
transfer $Q^2$, relative to the expectation in the standard model.  The
$e^+p$ center-of-mass energy has been $\sqrt{s}=300$ GeV.  Including the
new data presented recently at the 1997 Lepton-Photon Symposium
\cite{straub} \footnote{For a detailed discussion of the 1994-96 data
  see, for example, Ref.\ \cite{sirois}.}, H1 and ZEUS each observe 18
neutral current (NC) events at $Q^2 > 1.5 \cdot 10^4$ GeV$^2$, while H1
expects $8.0 \pm 1.2$ and ZEUS about 15 events.  At H1, the excess is
concentrated in the rather narrow mass range 187.5 GeV $\leq M =
\sqrt{xs} \leq$ 212.5 GeV where 8 events are observed with $1.53 \pm
0.29$ expected.  However, in the same region, ZEUS finds roughly the
expected number of events. Conversely, in the region $x>0.55$, $y =
Q^2/M^2 >0.25$ where ZEUS finds 5 events with $1.51 \pm 0.13$ expected,
H1 observes no excess. A surplus of events is also observed in charged
current (CC) scattering, although with smaller statistical significance.
At $Q^2 > 10^4$ GeV$^2$, H1 and ZEUS together find 28 events and expect
$17.7 \pm 4.3$.

The clustering of the H1 events at a fixed value of $M = \sqrt{xs}$
suggests the production of a resonance with leptoquark quantum numbers
and mass $M \simeq 200$ GeV. On the other hand, ZEUS has 4 events
clustered at a somewhat higher mass $M \simeq 225$ GeV.  Given the
experimental mass resolution of 5 and 9 GeV, respectively, it appears
unlikely that both signals come from a single narrow resonance
\cite{straub,BB}.  Rather, the excess may be a continuum effect
resulting from contact interactions.  Although the anomalous number of
events is not large enough to clearly exclude statistical fluctuations
as their origin, and the differences among the H1 and ZEUS data are
somewhat puzzling, it is important to investigate possible
interpretations within and beyond the standard model.

\section{Standard model and new physics}

Leaving aside the very small uncertainties in electroweak parameters and
radiative corrections \cite{H1data,Zdata}, the main theoretical
uncertainty on the high-$Q^2$ cross sections in the standard model comes
from structure functions.  The latter are obtained by extrapolation of
measurements at lower $Q^2$ using next-to-leading order evolution
equations.  For presently available parametrizations the HERA
collaborations have estimated this uncertainty to be about 7\,\%
\cite{H1data,Zdata}.  Attempts \cite{partons} to add to the conventional
parton densities a new valence component at very large $x$ but low
$Q^2$, and to feed down this enhancement to lower $x$ by evolution to
very high $Q^2$, fail to increase the cross sections by a sufficient
amount because of the constraints put by the fixed-target data.
Explanations based on a strong intrinsic charm component \cite{charm}
generated by some nonperturbative mechanism do not seem to be more
successful.  In fact, up to this day no standard model effect is known
which could explain the observed surplus of events.  Moreover, so far no
hint of a deviation from the perturbative evolution of structure
functions in QCD up to $Q^2 \simeq 10^4$ GeV$^2$ has been found in the
data.  Whatever mechanism is responsible for the HERA anomaly, it must
have quite a rapid onset.

Thus it is rather safe to conclude that either the excess is a
statistical fluctuation, or it is very likely produced by new physics
beyond the standard model.  This immediately raises the question whether
one is dealing with a (not necessarily single) resonance or with a
continuum effect.  Clearly, the most exciting speculation is the
existence of a new particle.  Being supposedly produced as a $s$-channel
resonance in $e^+q$ or $e^+ \bar q$ collisions, this new member of the
particle zoo must be a boson and carry simultaneously lepton and quark
quantum numbers.  Such species are generically called leptoquarks.  In
the following short overview, we focus on the leptoquark hypothesis
\footnote{See also the discussion by T. Rizzo in these proceedings, Ref.
  \cite{rizzo}.}.  Alternative interpretations of the HERA anomaly have
been discussed by other speakers at this Workshop. In particular, we
refer to the contribution to these proceedings by H.\ Dreiner on squarks
with $R$-parity violating couplings \cite{dreiner1}.  The case of
contact interactions is considered, for example, in Ref.
\cite{contacti}.

\section{Phenomenological framework}

\begin{table}[htp]
\begin{center}
\begin{tabular}{llllllll}
\hline \hline
\rule{0mm}{5mm}
$LQ$ & 
   $Q$ & 
     Decay & BR &
    Coupling & 
   Limits & HERA \\
&
  &
     Mode & $e^{\pm}\,j$ & $\lambda_{L,R}$
       & Ref.\ \cite{Leurer} & estimates \\[1mm]
\hline \rule{0mm}{5mm}
      &        &   $e_L u$   &     & \hspace{2.2mm}$g_L$  &  &  \\
$S_0$ & $-1/3$ &   $\nu_L d$ &  \raisebox{1.5ex}[-1.5ex]{$\frac{1}{2}$}  
      & $-g_L$ &   \raisebox{1.5ex}[-1.5ex]{$g_L < 0.06$} &
                                \raisebox{1.5ex}[-1.5ex]{0.40}  \\
      &        &   $e_R u$   & $1$ & \hspace{2.2mm}$g_R$  & $g_R < 0.1$
                                                            & 0.28
\\[2mm]
$\tilde{S}_0$ & 
  $-4/3 $      &
     $e_R d$ & 1 &
   \hspace{2.2mm}$g_R$ &
   $g_R<0.1$    & 0.30
\\[2mm]
& $+2/3$ & $\nu_L u$ & 0 & \hspace{2.2mm}$\sqrt{2}g_L$  &  & $-$ \\
&  & $\nu_L d$ &   & $-g_L$ &  & \\
\raisebox{1.5ex}[-1.5ex]{$S_1$} &
  \raisebox{1.5ex}[-1.5ex]{$-1/3$}  & 
     $e_L u$ &  
           \raisebox{1.5ex}[-1.5ex]{$\frac{1}{2}$} &
              $-g_L$ & 
   \raisebox{1.5ex}[-1.5ex]{$g_L<0.09$}    
      & \raisebox{1.5ex}[-1.5ex]{0.40} \\
& $-4/3$ & $e_L d$ & 1 & $-\sqrt{2}g_L$ &  & 0.21 
\\[2mm]
&  &  $\nu_L d$ & 0 & \hspace{2.2mm}$g_L$ & & $-$ \\
& \raisebox{1.5ex}[-1.5ex]{$-1/3$} & $e_R u$ & 1 & \hspace{2.2mm}$g_R$ & 
   \raisebox{1.5ex}[-1.5ex]{$g_L < 0.09$} & 0.30 \\
\raisebox{1.5ex}[-1.5ex]{$V_{1/2}$} & & $e_L d$ & &
\hspace{2.2mm}$g_L$ & & 0.32 \\ 
& \raisebox{1.5ex}[-1.5ex]{$-4/3$} & $e_R d$ & 
     \raisebox{1.5ex}[-1.5ex]{1} & \hspace{2.2mm}$g_R$ 
         & \raisebox{1.5ex}[-1.5ex]{$g_R < 0.05$} & 0.32
\\[2mm]
& $+2/3$ & $\nu_L u$ & 0 & \hspace{2.2mm}$g_L$ & & $-$ \\
\raisebox{1.5ex}[-1.5ex]{$\tilde{V}_{1/2}$} & $-1/3$ & 
     $e_L u $& 1 & \hspace{2.2mm}$g_L$ &
        \raisebox{1.5ex}[-1.5ex]{$g_L<0.09$} & 0.32
\\[3mm]
\hline \rule{0mm}{5mm}
&  &  $\nu_L \bar{u}$ & 0 & \hspace{2.2mm}$g_L$ & & $-$ \\
& \raisebox{1.5ex}[-1.5ex]{$-2/3$} & $e_R \bar{d}$ & 1 & $-g_R$ & 
   \raisebox{1.5ex}[-1.5ex]{$g_L < 0.1$} & 0.052 \\
\raisebox{1.5ex}[-1.5ex]{$S_{1/2}$} & & $e_L \bar{u}$ & &
\hspace{2.2mm}$g_L$ & & 0.026 \\ 
& \raisebox{1.5ex}[-1.5ex]{$-5/3$} & $e_R \bar{u}$ & 
     \raisebox{1.5ex}[-1.5ex]{1} & \hspace{2.2mm}$g_R$ & 
         \raisebox{1.5ex}[-1.5ex]{$g_R < 0.09$} & 0.026
\\[2mm]
& $+1/3$ & $\nu_L \bar{d}$ & 0 & \hspace{2.2mm}$g_L$ & & $-$ \\
\raisebox{1.5ex}[-1.5ex]{$\tilde{S}_{1/2}$} & $-2/3$ & 
     $e_L \bar{d} $& 1 & \hspace{2.2mm}$g_L$ &
        \raisebox{1.5ex}[-1.5ex]{$g_L<0.1$} & 0.052
\\[2mm]
      &        &   $e_L \bar{d}$ & & \hspace{2.2mm}$g_L$  &  &  \\
$V_0$ & $-2/3$ &   $\nu_L \bar{u}$ &  
                                 \raisebox{1.5ex}[-1.5ex]{$\frac{1}{2}$}  
 & \hspace{2.2mm}$g_L$ & \raisebox{1.5ex}[-1.5ex]{$g_L < 0.05$} & 
                                 \raisebox{1.5ex}[-1.5ex]{0.080} \\
& & $e_R \bar{d}$ & $1$ & \hspace{2.2mm}$g_R$  & $g_R < 0.09$ & 0.056
\\[2mm]
$\tilde{V}_0$ & $-5/3$ & $e_R \bar{u}$ & 1 &
    \hspace{2.2mm}$g_R$ & $g_R<0.09$& 0.027
\\[2mm]
& $+1/3$ & $\nu_L \bar{d}$ & 0 & \hspace{2.2mm}$\sqrt{2}g_L$ & & $-$ \\
&  & $\nu_L \bar{u}$ &   & \hspace{2.2mm}$g_L$ &  &  \\
\raisebox{1.5ex}[-1.5ex]{$V_1$} &
  \raisebox{1.5ex}[-1.5ex]{$-2/3$}  & 
     $e_L \bar{d}$ &  
           \raisebox{1.5ex}[-1.5ex]{$\frac{1}{2}$} & $-g_L$ & 
   \raisebox{1.5ex}[-1.5ex]{$g_L<0.04$}           & 
           \raisebox{1.5ex}[-1.5ex]{0.080} \\
& $-5/3$ & $e_L \bar{u}$ & 1 & \hspace{2.2mm}$\sqrt{2}g_L$ &  & 0.019 
\\[1mm]
\hline \hline
\end{tabular}
\caption{Scalar ($S$) and vector ($V$) leptoquarks, and their
  electric charges $Q$, decay modes, branching ratios into charged
  lepton + jet channels, and Yukawa couplings. Given are also the most
  stringent low-energy bounds and the couplings deduced from the 1994-96
  HERA data \protect\cite{krsz}.  Inclusion of the 1997 data decrease
  the couplings by about 15\%.  Using the H1 data alone would roughly
  give the couplings shown above.}
\label{tabprop}
\end{center}
\end{table}

Leptoquarks appear in extensions of the standard model involving
unification, technicolor, compositeness, or $R$-parity violating
supersymmetry. In addition to their couplings to the standard model
gauge bosons \footnote{The explicit form can be found, e.g., in Ref.
  \cite{gaugecoupl}.}, leptoquarks have Yukawa-type couplings to
lepton-quark pairs.  In the generally adopted framework described in
Ref.\ \cite{BRW}, the Yukawa couplings are taken to be dimensionless and
$SU(3)\times SU(2)\times U(1)$ symmetric.  Moreover, they are assumed to
conserve lepton and baryon number in order to avoid rapid proton decay,
to be non-zero only within one family in order to exclude FCNC processes
beyond CKM mixing, and chiral in order to escape the very strong bounds
from leptonic pion decays.

The allowed states can be classified according to spin, weak isospin and
fermion number. The nine possible scalar and vector leptoquarks are
listed in Tab.\ \ref{tabprop}.  We use the notation introduced in Ref.\ 
\cite{aachen} and generally employed in experimental papers: scalars are
denoted by $S_I$, vectors by $V_I$, $I$ being the weak isospin, and
isomultiplets with different hypercharges are distinguished by a tilde.
States in the upper half of Tab.\ \ref{tabprop} carry fermion number
$F=2$, those in the lower half have fermion number $F=0$.  Given are
also the electric charges, the decay modes for first generation
leptoquarks with the respective branching ratios, and the Yukawa
couplings generically called $\lambda_{L,R}$.  The indices $L,R$ refer
to the chirality of the lepton.  As a consequence of the assumption that
low-mass leptoquarks have either $L$- or $R$-couplings, but not both at
the same time, the branching fractions to a charged lepton final state
can only be 1, 0.5, or 0.

The Yukawa couplings of the leptoquark states summarized in Tab.\ 
\ref{tabprop} are given by the effective Lagrangeans \cite{BRW}
\begin{eqnarray}
\displaystyle
{\cal L}_{\rm eff}^S & = & \displaystyle 
  \left(g_L \bar{q}_L^c i\tau_2 l_L
+ g_R \bar{u}_R^c e_R \right) S_0
+ g_R \bar{d}_R^c e_R \tilde{S}_0
+ g_L \bar{q}_L^c i\tau_2\vec{\tau} l_L \vec{S}_1 \nonumber\\[1.5ex] 
 & &  \displaystyle
+ \left(g_L \bar{u}_R l_L
+ g_R \bar{q}_L i\tau_2 e_R \right) S_{1/2}
+ g_L \bar{d}_R l_L \tilde{S}_{1/2}~, \label{leffs}\\[3ex]
\displaystyle
{\cal L}_{\rm eff}^V & = & \displaystyle
  \left( g_L \bar{d}_R^c \gamma_{\mu} e_L
+ g_R \bar{q}_L^c \gamma_{\mu} e_R \right) V_{1/2}^{\mu}
+ g_L \bar{u}_R^c \gamma_{\mu} l_L \tilde{V}_{1/2}^{\mu}
\nonumber\\[1.5ex] & & \displaystyle
+ \left( g_L \bar{q}_L \gamma_{\mu} l_L 
+ g_R \bar{d}_R \gamma_{\mu} e_R \right) V_0^{\mu}
+ g_R \bar{u}_R \gamma_{\mu} e_R \tilde{V}_0^{\mu}
\label{leffv}\\[1.5ex] & & \displaystyle
+ g_L \bar{q}_L \vec{\tau} \gamma_{\mu} l_L \vec{V}_1^{\mu}~.
\nonumber
\end{eqnarray}
Here, $c$ denotes charge conjugation, $q_L$ and $l_L$ are the
left-handed quark and lepton weak isospin doublets, 
and $u_R$, $d_R$ and $e_R$ the right-handed singlets.

\section{Production and decay}

With the above couplings the resonance cross section in $ep$ scattering
is given by
\begin{equation}
\sigma = N_\sigma \frac{\pi}{4s} \lambda_{L,R}^2 q_f(M^2/s,\mu^2)~,
\end{equation}
where $q_f(x,\mu^2)$ is the density of quarks (or antiquarks) with
flavour $f$ in the proton, and $N_\sigma = 1\,(2)$ for scalars
(vectors).  The relevant scale $\mu$ is expected to be of order of the
leptoquark mass $M$.  The coupling constant $\lambda_{L,R}$ can be read
off from Tab.\ \ref{tabprop}. Obviously, leptoquarks with fermion number
$F=0\,(2)$ can be produced from valence quarks in $e^+q\,(e^-q)$ fusion.
This is essential for the interpretation of the HERA anomaly: the
coupling strength required for $F=0$ resonance production is much
smaller than the one for $F=2$ production.

Having only couplings to standard model particles, leptoquarks decay
exclusively to lepton-quark pairs. The partial width per channel is
given by
\begin{equation}
\Gamma = \frac{N_\Gamma}{16\pi} \lambda_{L,R}^2 M = 
350\,{\rm MeV}\,N_\Gamma
\left(\frac{\lambda}{e}\right)^2 \left(\frac{M}{200\,{\rm GeV}}\right)~, 
\label{width}
\end{equation}
$N_\Gamma$ being 1 for scalars and 2/3 for vectors. Hence, leptoquarks
are very narrow for masses in the range accessible at HERA, and for
couplings weaker than the electromagnetic coupling strength
$e=\sqrt{4\pi\alpha}$.  Obviously, only states with charge 2/3 can be
produced in $e^+q$ fusion and subsequently decay into $\bar{\nu}_e q$.
For chiral couplings $\lambda_L \not= 0$ and $\lambda_R = 0$, this
leaves only the vector leptoquarks $V_0$ and $V_1$ as possible sources
of CC final states. Similarly, in $e^+\bar{q}$ fusion only the charge
1/3 scalar leptoquarks $S_0$ and $S_1$ can give rise to CC events.  The
branching fractions into a charged lepton plus jet and neutrino plus jet
are 50\% each.  This is a second feature which plays an important role
in interpretations of the HERA data.

In order to explain the observed excess of high-$Q^2$ events at HERA by
the production and decay of a 200 GeV leptoquark, one roughly needs
$\lambda_{L,R} \simeq e$ for $F=2$ states and $\lambda_{L,R} \simeq
e/10$ for $F=0$.  The factor 10 difference in $\lambda$ simply reflects
the factor 100 difference in the sea and valence quark densities in the
region of $x$ and $Q^2$ where the signal resides.  Similarly, the
coupling of $F=0$ leptoquarks to the $d$ quark has to be two times
larger than the coupling to the $u$ quark in order to compensate the
factor four difference in the corresponding quark densities. These
simple rules of thumb describe the main pattern in the couplings found
in detailed analyses \cite{krsz,LQ}, and shown in the last column of
Tab.\ \ref{tabprop}.

\begin{figure}[htbp] 
\unitlength 1mm
\begin{picture}(114,95)
\put(15,0){
\epsfig{
file=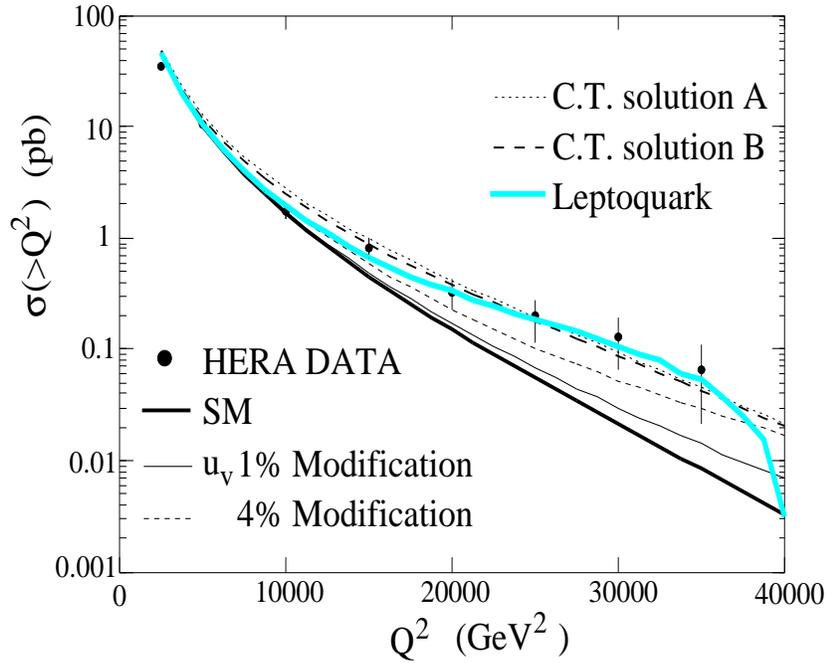,
height=9cm,
width=11cm,
bbllx=35,bblly=195,bburx=465,bbury=510,clip
}
}
\end{picture}
\caption{Cross section integrated above a given minimum $Q^2$ 
  in the standard model and in the presence of the $S_{1/2}^L$
  leptoquark with $M=200$ GeV and $\lambda_L = 0.025$ 
  compared with the 1994 - 96 data. 
  Also shown are effects due to an enhancement of the valence
  quark density and due to contact interactions. 
  (From \protect\cite{cao}.)}
\label{fig1}
\end{figure}

Figure \ref{fig1} (taken from Ref.\ \cite{cao}) shows the $e^+p$ cross
section integrated above a given minimum value of $Q^2$ in a scenario
with a 200 GeV $S_{1/2}$ leptoquark in comparison with the 1994 - 96
data \cite{H1data,Zdata} and the standard model expectation.  The Yukawa
coupling $\lambda_L$ is taken to be 0.025 in conformity with the
estimate given in Tab.\ \ref{tabprop}.  As one can see, the leptoquark
hypothesis has provided quite a satisfactory interpretation of the data
from the 1994 - 96 runs.
 
\begin{figure}[hbt] 
\unitlength 1mm
\begin{picture}(114,120)
\put(15,0){
\epsfig{
file=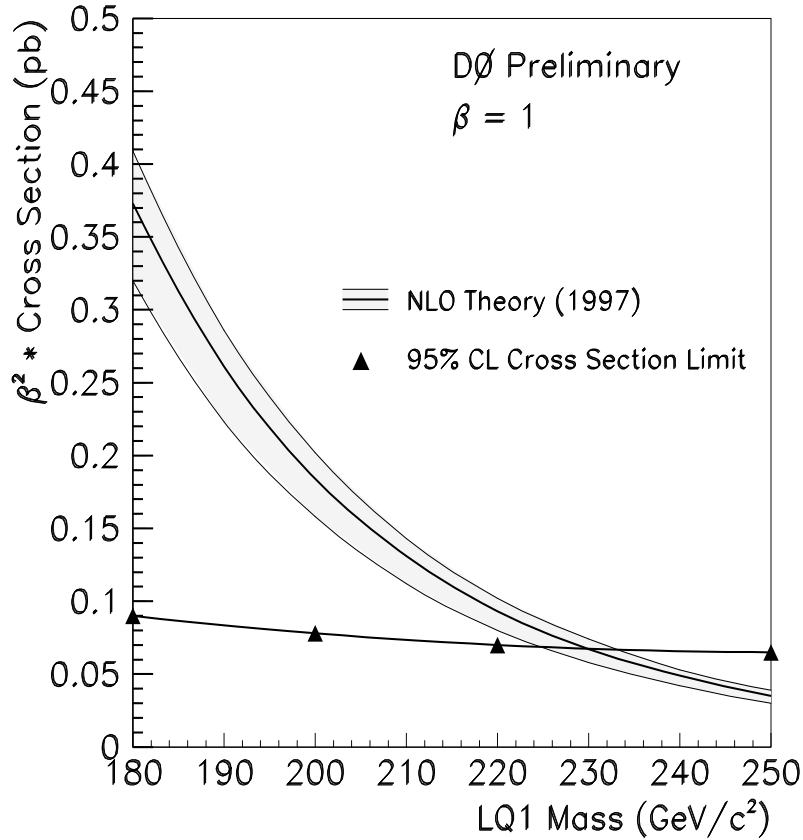,
height=15cm,
width=11cm,
bbllx=150,bblly=460,bburx=550,bbury=900,clip
}
}
\end{picture}
\caption{The D0 95\,\% CL limit on the production cross section times
  branching ratio into the $eejj$ channel ($\beta = B_{eq}$) 
  for first generation leptoquarks.  
  The band shows the NLO theoretical prediction \protect\cite{kraemer}.
  (From \protect\cite{D0}.)}
\label{fig2}
\end{figure}

\section{Bounds}

The leptoquark masses and couplings are constrained by a number of low-
and high-energy experiments.  Direct searches for leptoquarks have been
performed at the Tevatron, at HERA and at LEP.  Recently \footnote{See
  also the discussions by S. Eno and J. Conway in these proceedings,
  Ref. \cite{tev}.}, both collaborations CDF and D0 have improved their
mass limits for scalar leptoquarks considerably.  D0 excludes first
generation leptoquarks with masses below 225 GeV assuming a branching
ratio $B_{eq}=1$ for decays into $e^{\pm}$ and a jet \cite{D0}, whereas
CDF quotes a limit of 213 GeV \cite{cdf} (all mass limits are at 95\,\%
CL). For branching ratios less than one, the limits are weaker, e.g., $M
> 176$ GeV for $B_{eq}=0.5$ \cite{D0}.  Figure \ref{fig2} shows the D0
limit on the production cross section times the branching ratio
$\beta^2$ for the search channel $eejj$, $\beta$ being the branching
fraction $B_{eq}$.  Even stronger bounds hold for vector leptoquarks:
298 GeV for $B_{eq}=1$ and 270 GeV for $B_{eq}=0.5$ \cite{vector}.  The
corresponding mass limits on second and third generation scalar
leptoquarks are $M > 184$ GeV for $B_{\mu q} = 1$ and $M > 98$ GeV for
$B_{\tau q} = 1$, respectively \cite{lqst}.  The above constraints
follow from pair production mainly by $q\bar q$ annihilation, and are
therefore practically independent of the unknown Yukawa coupling
$\lambda$.

\begin{figure}[htbp] 
\unitlength 1mm
\begin{picture}(114,155)
\put(-15,-54){
\epsfxsize=18cm
\epsfysize=25cm
\epsfbox{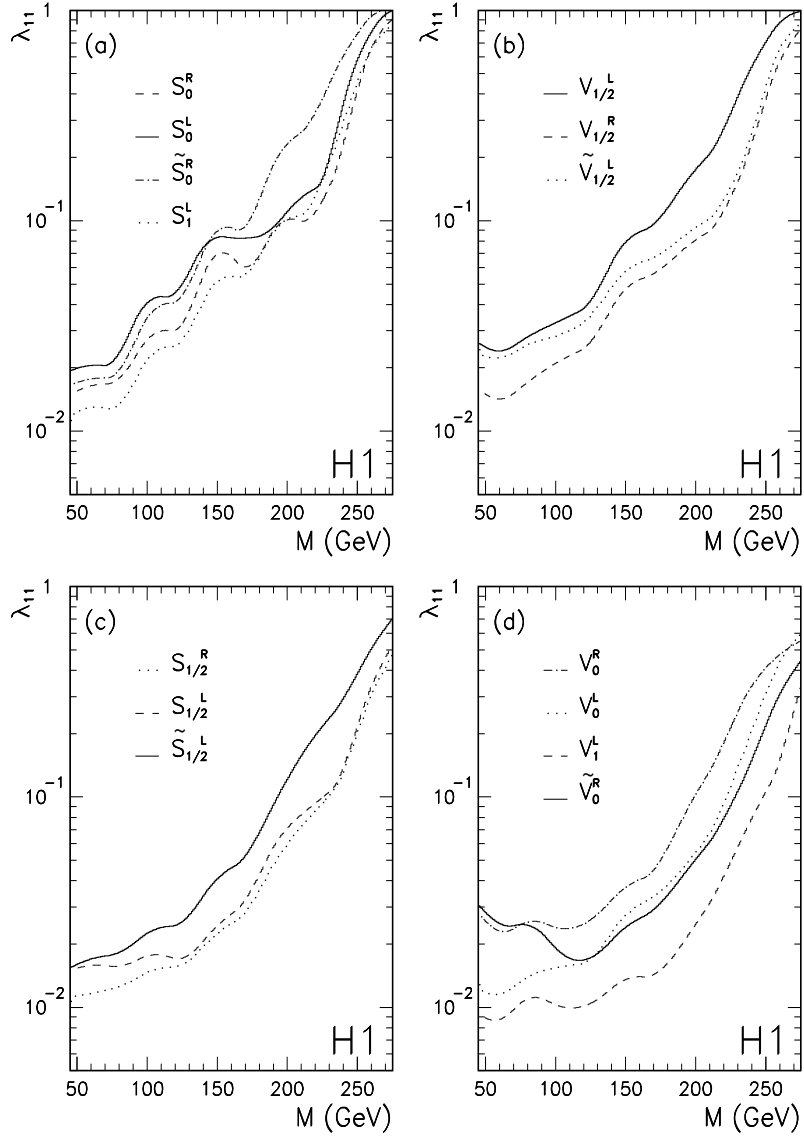}
}
\end{picture}
\caption{H1 upper limits at 95\,\% CL from $e^-p$ and $e^+p$ combined 
  for scalar and vector leptoquarks with fermion number 2 (a,b) and 0
  (c, d). The upper indices $L,R$ refer to the models $\lambda =
  \lambda_L$, $\lambda_R = 0$ and $\lambda = \lambda_R$, $\lambda_L = 0$,
  respectively.  The limits on $\lambda_L$ for $S_0$,
  $S_1$, $V_0$ and $V_1$ result from $e+X$ and $\nu+X$ final states.
  (From \protect\cite{h1lq}.)}
\label{fig3}
\end{figure}

In contrast, the mass bounds obtained at HERA \cite{h1lq,zeuslq} depend
on $\lambda$ and the quantum numbers specified in Tab.\ \ref{tabprop}.
For $\lambda = e$ and $F=0$ ($F=2$) leptoquarks the upper limits from
resonance production in $e^-p$ (0.4 pb$^{-1}$) and $e^+p$ (2.8
pb$^{-1}$) collisions at H1 reach up to 270 (245) GeV.  The
corresponding reach for $\lambda = 0.03$ is 170 (130) GeV, except in the
case of $V_0^R$ which is excluded up to $M=210$ GeV.  Figure \ref{fig3}
shows the detailed H1 bounds in the $(\lambda, M)$--plane for all
leptoquark species listed in Tab.\ \ref{tabprop}.  Heavy leptoquarks
generate effective contact interactions \cite{lqcontact} and can
therefore be probed by a general contact term analysis.  The outcome of
such a test is presented in Ref. \cite{h194}.

At LEP2, the most stringent but again $\lambda$-dependent mass bound
comes from the search for single-leptoquark production at
$\sqrt{s}=161$ and 172 GeV, and excludes masses for scalars with
$|Q|=5/3$ and 1/3 below 131 GeV assuming $\lambda \geq e$ \cite{leps}.
The upper limits on leptoquark masses from pair production \cite{lepp},
being close to half of the center of mass energy $\sqrt{s}$, are weaker
than the above limit, and also way below the Tevatron bounds.  Indirect
constraints from $t$- and $u$-channel exchange of leptoquarks in $e^+e^-
\to q \bar{q}$ are approaching an interesting sensitivity. From the very
recent analysis by OPAL \cite{eeqq} for $\sqrt{s}=130$ to 172 GeV we
infer upper limits on $\lambda$ between 0.2 and 0.7 assuming $M=200$
GeV.  We come back to this interesting search for virtual effects in
section 7.  In addition, similarly as at HERA, bounds on contact
interactions can be translated into constraints on heavy leptoquarks.
States with integer isospin $I=0$ and $I=1$ generate equal-helicity $LL$
and $RR$ contact terms, while leptoquarks with $I=1/2$ give rise to
opposite-helicity $RL$ and $LR$ contact terms.  These rules may also be
cast into the standard form of the effective Lagrangian~\cite{effl}
\begin{eqnarray}
{\cal L}_{\rm eff} & = & \sum_{i,k=L,R}\frac{ g^2_i}{M^2}\; \alpha_{ik}^q\;
(\bar{e}_i\gamma^{\mu} e_i)\,(\bar{q}_k\gamma_{\mu} q_k)
\nonumber\\ 
 & := & \sum_{i,k=L,R}\eta_{ik}\;\frac{ 4\pi}{\Lambda^2_{ik}}\;
(\bar{e}_i\gamma^{\mu} e_i)\,(\bar{q}_k\gamma_{\mu} q_k)~.
\label{Leff} 
\end{eqnarray}
The coefficients $\alpha_{ik}^q$ for $u \bar u$ and $d \bar d$ final
states are listed in Tab.\ \ref{tabcon1} \cite{krsz}.  Denoting the
signs of $\alpha_{ik}^q$ by $\eta_{ik}$, the scales $\Lambda_{ik}$ of
the contact interactions are related to the individual masses and
couplings of the leptoquarks by $\Lambda^2_{ik}=4\pi M^2/g^2_i
|\alpha_{ik}^q|$.

\newcommand{\ft}{$\phantom{-}\frac{1}{2}$}
\newcommand{\fm}{$-\frac{1}{2}$}
\newcommand{\mo}{$\phantom{-}1$}
\begin{table}[ht]
\begin{center}
\begin{tabular}{lllllllll}
\hline \hline \rule{0mm}{5mm}
&\multicolumn{4}{c}{\rule{0mm}{5mm}$u \bar u$ final state}&
\multicolumn{4}{c}{$d \bar d$ final state}\\[1mm]
\cline{2-9} \rule{0mm}{5mm}
$\alpha_{ik}^q$          &$RR$ &$LL$ &$RL$ &$LR$ &$RR$ &$LL$ &$RL$ &$LR$ 
\\[1mm]
\hline \rule{0mm}{5mm}
$S_0$         & \ft & \ft &     &     &     &     &     &\\[1mm]
$\tilde{S}_0$    &     &     &     &     &\ft  &     &     & \\[1mm] 
$S_1$         &     & \ft &     &     &     & \mo &     & \\[1mm]
$V_{1/2}$     &     &     & \mo &     &     &     & \mo & \mo  \\[1mm]
$\tilde{V}_{1/2}$&     &     &     & \mo &     &     &     & \\[1mm]
\hline \rule{0mm}{5mm}
$S_{1/2}$     &     &     & \fm & \fm &     &     & \fm & \\[1mm]
$\tilde{S}_{1/2}$&     &     &     &     &     &     &     & \fm \\[1mm]
$V_0$         &     &     &     &     &$-1$ & $-1$&     & \\[1mm]
$\tilde{V}_0$    &$-1$ &     &     &     &     &     &     & \\[1mm]
$V_1$         &     &$-2$ &     &     &     &$-1$ &     & \\[1mm]
\hline \hline
\end{tabular}
\caption{The coefficients $\alpha_{ik}^q$ in the Lagrangian Eq.\
  \ref{Leff} for contact interactions generated by heavy leptoquark
  exchange, and in the corresponding helicity amplitudes Eqs.\ 
  \ref{helf0} and \ref{helf1} \protect\cite{krsz}.}
\label{tabcon1}
\end{center}
\end{table}

Finally, indirect bounds on Yukawa couplings and masses can also be
derived from low-energy data \cite{Leurer}.  The most restrictive bounds
come from atomic parity violation and lepton and quark universality, at
least for first generation leptoquarks and chiral couplings. The maximum
allowed couplings for $M=200$ GeV are given in Tab.\ \ref{tabprop}
\cite{krsz}.

\section{Difficulties and remedies}

Whereas the coupling strength $\lambda$ required for $F=0$ leptoquarks
to explain the observed excess of events is compatible with all existing
bounds, the coupling necessary for $F=2$ leptoquarks is already excluded
by the low-energy constraints, and also at the borderline of getting in
conflict with LEP2 data.  Moreover, with such strong couplings, $F=2$
leptoquarks should have shown up in $e^-p$ scattering at HERA
\cite{h1lq,zeuslq}, despite of the low luminosity of the previous $e^-p$
run. The point is that in $e^-p$ the $F=2$ states can be produced off
the valence quark component of the proton.  As can be seen from Figs.\ 
3a and b, the existing HERA bounds indeed rule out an interpretation of
the high-$Q^2$ anomaly in terms of $F=2$ leptoquarks with $M \simeq 200$
GeV.

Furthermore, since vector leptoquarks cannot be responsible for an
excess of events at $M \simeq 200$ to 225 GeV because of the high
Tevatron mass bounds, only the two scalar doublets $S_{1/2}$ and
$\tilde{S}_{1/2}$ remain from the whole Tab.\ \ref{tabprop} as a
possible source of the signal. However, also these solutions have
difficulties. Firstly, the Tevatron mass limits require scalar
leptoquarks of the first generation with $M \simeq 200$ GeV to have
branching ratios into $e\,+\,jet$ final states less than about
\footnote{This value follows from the D0 limit \cite{D0} alone. An even
  smaller branching ratio is required by the combined D0 and CDF
  bounds.} 0.7, whereas in the framework considered in Tab.\
\ref{tabprop}, $S_{1/2}$ and $\tilde{S}_{1/2}$ are expected to have
$B_{eq}=1$ (or 0, but then they cannot be produced in $e^+p$).
Secondly, the scalar doublets do not give rise to CC events. As already
mentioned, among the $F=0$ leptoquarks only the vector states $V_0$ and
$V_1$ decay into $\bar{\nu}_e + jet$. However, vector leptoquarks are
excluded.  Thirdly, any single-resonance interpretation of the
high-$Q^2$ events has difficulties to explain the distributions in $M$
or $x$ simultaneously for H1 and ZEUS.

Thus it seems that the leptoquark interpretation of the HERA high-$Q^2$
events points at quite complicated scenarios involving more than just a
single leptoquark at a time, and different couplings, not just the
coupling to first generation fermions with given chirality. The task is
clear: find a model which predicts a sufficiently small branching
fraction for $S \to e q$, to wit $B_{eq} < 0.5$, a sufficiently large
branching into $S \to \nu q$, and a broad mass bump rather than a narrow
resonance signature.  Several possibilities have been suggested:
$SU(2)\times U(1)$ violating, intergenerational couplings \cite{ks,agm}
and leptoquark mixing \cite{babu}, LQ models with additional vector-like
fermions \cite{hr,rizzo}, and squarks with $R$-parity violating
couplings \cite{squarks,dreiner1}.

The latter proposition is clearly the most interesting one, since it can
be realized in a supersymmetric extension of the standard model which is
welcome for many other reasons.  In the minimal supersymmetric standard
model, one can have a renormalizable, gauge invariant operator in the
superpotential that violates $R$-parity conservation and couples squarks
to quarks and leptons. In such models, squarks act as leptoquarks.  More
precisely, direct couplings to lepton-quark pairs exist for the singlets
$\tilde{d}_R^n$ and the doublets ($\bar{\tilde{d}}_L^n$,
$\bar{\tilde{u}}_L^n$), $n$ being the generation index.  The quantum
number assignment for these squarks is identical to the assignment for
the states $S_0$ and $\tilde{S}_{1/2}$, respectively, given in Tab.\
\ref{tabprop}. Consequently, much of what has been said about leptoquark
production and virtual exchange can be carried over to the squark
scenario. In particular, squarks can be resonance-produced at HERA
\cite{squarks,rp}:
\begin{equation}
e^+ d_R^n \rightarrow \tilde{u}_L^m
~~~~~~ (\tilde{u}^m = \tilde{u}, \tilde{c}, \tilde{t}),
\end{equation}
\begin{equation}
e^+ \bar{u}_L^m \rightarrow \bar{\tilde{d}}_R^n
~~~~~~ (\tilde{d}^n = \tilde{d}, \tilde{s}, \tilde{b}).
\end{equation}
However, because of the usual $R$-parity conserving interactions one
naturally expects the branching ratio for $\tilde{q} \to e\,+\,jet$ to
be smaller than unity.  In addition, one can get CC-like final states,
e.g., through the decay chain $\tilde{q} \to q \chi \to q \nu + \cdots$,
$\chi$ being either a neutralino or chargino.  Thus it appears possible
to avoid two of the main problems encountered in the generic leptoquark
models.  Taking into account the relevant experimental constraints on
masses and Yukawa couplings \cite{dreiner2}, one finds three allowed
channels:
\begin{equation}
e^+d_R \to \tilde{c}_L ,~~ e^+d_R \to \tilde{t}_L ,~~ e^+s_R \to
\tilde{t}_L. 
\label{sqchannel}
\end{equation}
In the MSSM each fermion has two superpartners, $\tilde{f}_L$ and
$\tilde{f}_R$, which mix in general. In the case of stop this mixing may
be sizeable and lead to two mass eigenstates with a small but pronounced
mass difference. In this way the difficulty to interpret the excess of
events as a single-resonance effect may also find a reasonable solution
\cite{kon}.  For a detailed review of the squark interpretation of the
HERA high-$Q^2$ events we refer to the contribution by H. Dreiner
\cite{dreiner1}.


\section{Related predictions}

Leptoquarks (squarks) which couple in the $s$-channel in $e q \to e q$
contribute via $t/u$-channel exchange to the crossed reactions $e^+e^-
\to q \bar q$ \cite{krsz} and $q \bar q \to e^+e^-$ \cite{pp2}.
Therefore, the leptoquark (squark) hypothesis can in principle be tested
at LEP2 and at the Tevatron probing the same Yukawa couplings as at
HERA.  Effects on Drell-Yan production in hadronic collisions have been
studied in Ref. \cite{pp2} with emphasis on the experimental prospects
at the LHC.

The angular distribution and integrated cross section for hadron
production in $e^+e^-$ annihilation in the presence of any of the
leptoquarks of Tab.\ \ref{tabprop} can be found in Ref. \cite{krsz}.
Here, we give the results under the assumption that only a single state
contributes:
\begin{eqnarray}
\frac{\mbox{d}\sigma}{\mbox{d}\cos\theta}(e^+e^-\rightarrow q\bar{q}) 
& = &
\frac{3}{32\pi s} \Bigl\{
\left(|f_{RR}|^2 + |f_{LL}|^2\right) u^2 + 
\\ \nonumber & & \hspace{10.5mm}
\left(|f_{RL}|^2 + |f_{LR}|^2\right) t^2
\Bigr\}~, 
\label{diff-xsec} 
\end{eqnarray}
where
\begin{equation}
f_{ik} = \frac{Q^{eq}_{ik}}{s} - \frac{g^2_i \alpha_{ik}^{q}}{t-M^2} 
\label{helf0} 
\end{equation}
for $F=0$ leptoquarks
(or the squark doublets ($\bar{\tilde{d}}_L^n$, $\bar{\tilde{u}}_L^n$)), and   
\begin{equation}
f_{ik} = \frac{Q^{eq}_{ik}}{s} - \frac{g^2_i \alpha_{ik}^{q}}{u-M^2} 
\label{helf1} 
\end{equation}
for $F=2$ leptoquarks (or the squark singlets $\tilde{d}_R^n$).  The
indices $i,k=R,L$ refer to the handedness of the electron and quark in
the process $e^-_ie^+\to q_k\bar{q}$.  The generalized charges in the
standard $\gamma,Z$ exchange amplitudes have been abbreviated by
$Q^{eq}_{ik}$ where
\begin{equation}
Q^{eq}_{ik}= e^2\,Q_eQ_q +\frac{g^e_ig^q_k}{1-m_Z^2/s} 
\end{equation}
with the left/right $Z$ charges of the fermions defined as 
\begin{eqnarray}
g^f_L&=&\frac{e}{s_Wc_W}\left[I^f_3-s^2_W Q_f\right]~,
\nonumber \\ 
g^f_R&=&\frac{e}{s_Wc_W}\left[ {} -s^2_W Q_f\right]
\nonumber
\end{eqnarray}
and $s_W=\sin\Theta_W$, $c_W=\cos\Theta_W$.  The Mandelstam variables
$t,u$ can be expressed by the production angle $\theta$: $t = -s(1 -
\cos\theta)/2$, $u = -s(1 + \cos\theta)/2$; they are both negative so
that the amplitudes for $LQ$ exchange do not change the sign when
$\theta$ is varied from the forward to the backward direction.
Simiarly, as in the case of contact interaction, Eq.\ \ref{Leff} and
Tab.\ \ref{tabcon1}, leptoquarks with integer isospin contribute to
equal-helicity $LL$ and $RR$ amplitudes, while leptoquarks with $I=1/2$
contribute to opposite-helicity amplitudes $RL$ and $LR$.

\begin{figure}[htbp] 
\unitlength 1mm
\begin{picture}(120,116)
\put(13.5,0){
\epsfxsize=11.5cm
\epsfysize=11cm
\epsfbox{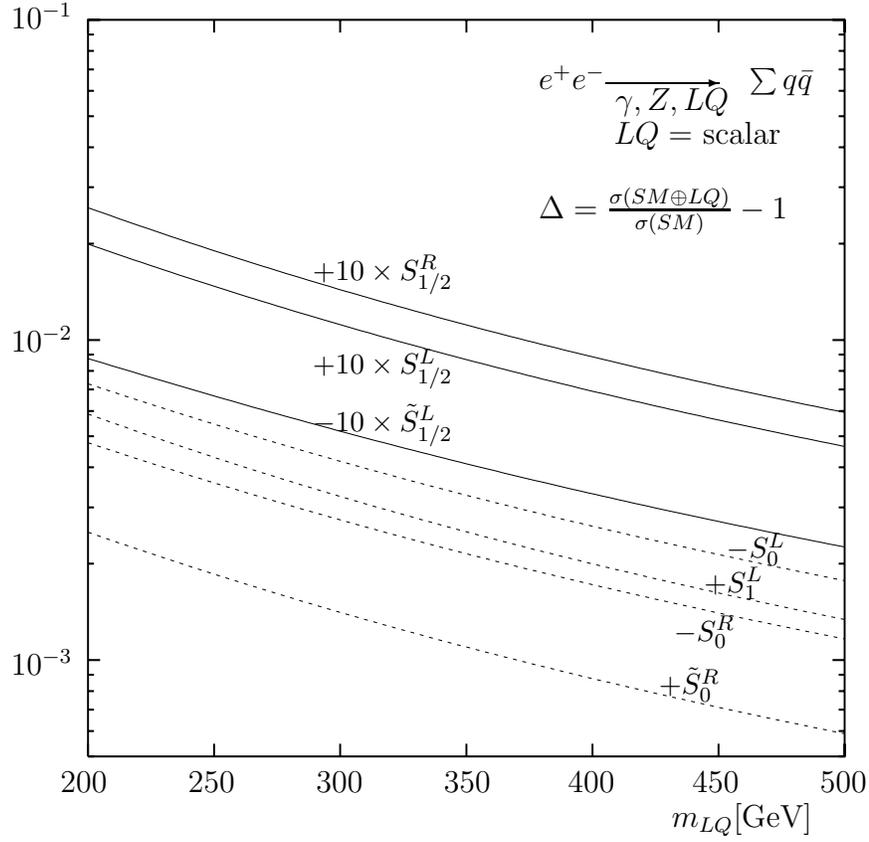}}
\setlength{\unitlength}{1mm}
\put(85,82){\makebox(0,0)[l]{$\Delta = \frac{\sigma(SM\oplus 
LQ)} {\sigma(SM)}-1$}}
\put(95,92){\makebox(0,0)[l]{$LQ = $ scalar}}
\put(85,98){\makebox(0,0)[l]{{$e^+e^-$} 
\raisebox{-1.5ex}{$\stackrel{}{\footnotesize \gamma, Z, LQ}$} 
~{$\sum q\bar{q}$}}}
\put(94,99){\vector(1,0){15}}
\put(55,53.5){\makebox(0,0)[l]{\small $-10\times \tilde{S}_{1/2}^L$}}
\put(55,61){\makebox(0,0)[l]{\small $+10\times S_{1/2}^L$}}
\put(55,73.5){\makebox(0,0)[l]{\small $+10\times S_{1/2}^R$}}
\put(101,19){\makebox(0,0)[l]{\small $+\tilde{S}_0^R$}}
\put(103,26){\makebox(0,0)[l]{\small $-S_0^R$}}
\put(107,32.5){\makebox(0,0)[l]{\small $+S_1^L$}}
\put(110,37){\makebox(0,0)[l]{\small $-S_0^L$}}
\put(112,1){\makebox(0,0){$m_{LQ}$[GeV]}}
\put(125.5,6){\makebox(0,0){500}}
\put(108.5,6){\makebox(0,0){450}}
\put(92,6){\makebox(0,0){400}}
\put(75.5,6){\makebox(0,0){350}}
\put(58.5,6){\makebox(0,0){300}}
\put(41.5,6){\makebox(0,0){250}}
\put(24.5,6){\makebox(0,0){200}}
\put(23,108){\makebox(0,0)[r]{$10^{-1}$}}
\put(23,65){\makebox(0,0)[r]{$10^{-2}$}}
\put(23,22){\makebox(0,0)[r]{$10^{-3}$}}
\end{picture}
\caption{Effect of $t/u$-channel exchange of scalar leptoquarks on the
  total hadronic cross section at LEP2 for $\protect\sqrt{s} = 192$ GeV
  \protect\cite{krsz}. The couplings have been fixed arbitrarily to
  $(g_L,g_R) = (0.1,0)$ or $(0,0.1)$ indicated by the upper indices
  ${L,R}$, respectively.}
\label{figxss}
\end{figure}

\begin{figure}[htbp] 
\unitlength 1mm
\begin{picture}(114,154)
\put(15,0){
\epsfig{
file=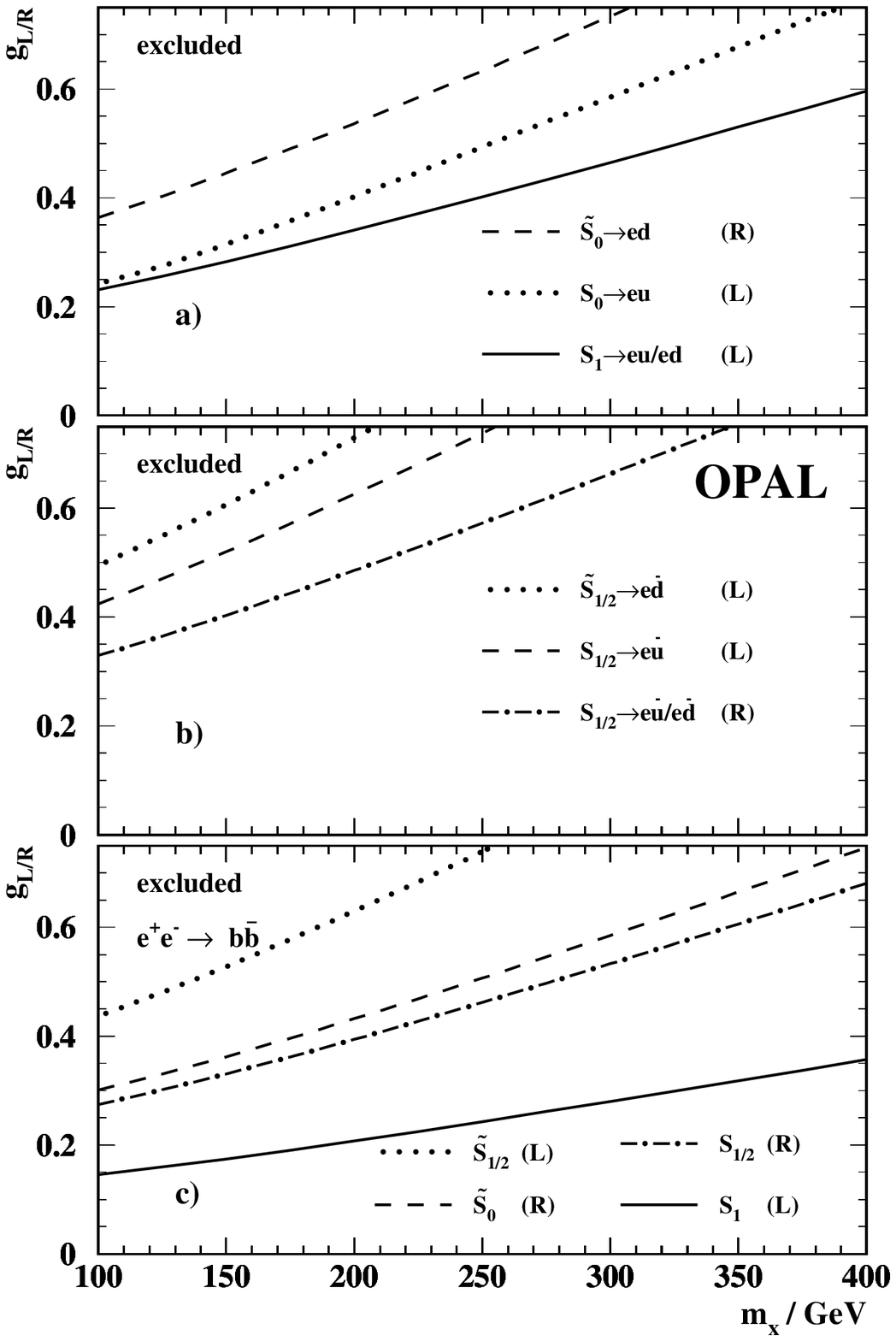,
height=15cm,
width=11cm,
bbllx=115,bblly=185,bburx=475,bbury=720,clip
}
}
\end{picture}
\caption{OPAL 95\,\% confidence exclusion limits on $g_{L,R}$ as a
  function of the mass $m_X$ for scalar leptoquarks: (a) and (b) are
  derived from $e^+e^- \to \sum q\bar{q}$, (c) from $e^+e^- \to
  b\bar{b}$. Excluded are the regions above the curves. (From
  \protect\cite{eeqq}.)}
\label{fig4}
\end{figure}

For small Yukawa couplings, the leptoquark contributions are dominated
by the interference with the standard model amplitudes leading to an
enhancement or suppression of the integrated cross section relative to
the standard model expectation.  Furthermore, the effects scale with
$g_{L,R}^2$ and, for large LQ-masses, with $1/M^2$.  For scalar
leptoquarks with couplings $g_{L,R}=0.1$, Fig.\ \ref{figxss} shows the
relative change of the total cross section of hadron production at LEP2
for $\sqrt{s} = 192$ GeV \cite{krsz}. Similar results are obtained for
vector states.  Generally, the effect is smaller for scalar leptoquarks
than for vectors, and for isopin doublets than for singlets and
triplets. It is interesting to note that the species $S_{1/2}$ and
$\tilde{S}_{1/2}$ which play a particularly important role in the
interpretation of the HERA high-$Q^2$ events, give the smallest effect
of all leptoquarks at LEP2, to wit $|\Delta| = O(10^{-4})$ for $g_L
\simeq 0.03$ and $M=200$ GeV.  The impact on cross sections for
individual quark flavours is of course somewhat stronger. On the other
hand, if the excess of events at HERA was due to the production of a
$F=2$ leptoquark in positron-antiquark scattering, the effect of
leptoquark exchange implied at LEP2 should be observable since the
Yukawa couplings would be larger in this case, $g_{L,R} \simeq 0.3$.
This is also suggested by the recent OPAL analysis \cite{eeqq} for
$\sqrt{s} = 130$ to 172 GeV shown in Fig.\ \ref{fig4}, which is already
sensitive to the $F=2$ scalars $S_0$, $\tilde{S}_0$ and $S_1$ for $M
\simeq200$ GeV and $g_{L,R}$ between about 0.3 and 0.5.

\section{Conclusions}

For the time being, it is an open question whether or not the excess of
high-$Q^2$ events observed at HERA is a statistical fluctuation or a
physical effect. If it is a real signal, then it very likely originates
from new physics beyond the standard model. Making this assumption, the
present data slightly favour a continuum mechanism, but do not yet allow
to exclude a resonance effect.  Both kinds of interpretations are
tightly constrained by measurements at LEP2 and the Tevatron, as well as
by low-energy data.  These bounds rule out the simplest leptoquark
scenarios, but leave some room for extended leptoquark models and, most
importantly, for a squark interpretation. Particularly difficult is the
explanation of anomalous CC events.  At any rate, if the excess of
high-$Q^2$ events is confirmed by future data related signals are likely
to show up soon in other experiments.


\vspace{10mm}
\noindent{\bf Acknowledgments}

\vspace{2mm}

\noindent
We wish to thank J.\ Kalinowski and P.\ Zerwas for the fruitful
collaboration on the subject of this talk. This work was supported
by the Bundesministerium f\"ur Bildung, Wissenschaft, Forschung und
Technologie, Bonn, Germany, Contracts 05 7BI92P (9) and 05 7WZ91P (0).




\vspace{7mm}
\noindent{\bf References}

\noindent

\begin{thebibliography}{99}

\bibitem{H1data} Adloff C \etal (H1 Collaboration) 1997 {\it Z.\ Phys.}   
  C {\bf 74} 191

\bibitem{Zdata} Breitweg J \etal (ZEUS Collaboration) 1997 {\it Z.\
    Phys.} C {\bf 74} 207

\bibitem{straub} Straub B 1997 {\it Talk presented at the XVIII
    International Symposium on Lepton Photon Interactions} (Hamburg)  

\bibitem{sirois} Sirois Y 1997 {\it These proceedings}

\bibitem{BB} Bassler U and Bernardi G 1997 {\it Z.\ Phys.} C {\bf 76}
  223 

\bibitem{partons} Kuhlmann S, Lai H L and Tung W K 1997 {\it
    hep-ph/9704338};  
  \newref Babu K S, Kolda C and March-Russell J 1997 {\it Phys.\ Lett.}
  {\bf B408} 268

\bibitem{charm} Gunion J F and Vogt R 1997 {\it hep-ph/9706252}; 
  \newref Melnitchouk W and Thomas A W 1997 {\it hep-ph/9707387}

\bibitem{rizzo} Rizzo T 1997 {\it These proceedings}

\bibitem{dreiner1} Dreiner H 1997 {\it These proceedings}

\bibitem{contacti} Babu K S \etal 1997 {\it Phys.\ Lett.} {\bf B402} 367;
  \newref Barger V \etal 1997 {\it Phys.\ Lett.} {\bf B404} 147;   
  {\it hep-ph/9707412};
  \newref Di Bartolomeo N and Fabbrichesi M 1997 {\it Phys.\ Lett.} 
  {\bf B406} 237;
  \newref Deshpande N G \etal 1997 {\it Phys.\ Lett.} {\bf B408} 288;
  \newref Cornet F and Rico J 1997 {\it hep-ph/9707299}

\bibitem{gaugecoupl} Bl\"umlein J and R\"uckl R 1993 {\it Phys.\
  Lett.} {\bf B304} 337; \newref
  Bl\"umlein J, Boos E and Kryukov A 1997 {\it Z.\ Phys.} C {\bf 76}
  137; {\it Phys.\ Lett.} {\bf B392} 150

\bibitem{BRW} Buchm\"uller W, R\"uckl R and Wyler D 1987 {\it Phys.\
  Lett.} {\bf B191} 442

\bibitem{aachen} Djouadi A, K\"ohler T, Spira M and Tutas J 1990
  {\it Z.\ Phys.} C {\bf 46} 679

\bibitem{Leurer} Leurer M 1994 {\it Phys.\ Rev.} D {\bf 49} 333;
   {\it Phys.\ Rev.} D {\bf 50} 536; \newref
   Davidson S, Bailey D and Campbell B 1994 {\it Z.\ Phys.} C {\bf 61} 
   613

\bibitem{krsz} Kalinowski J, R\"uckl R, Spiesberger H and Zerwas P M
  1997 {\it Z.\ Phys.} C {\bf 74} 595

\bibitem{LQ} Altarelli G, Ellis J, Giudice G F, Lola S and Mangano M L
   1997 {\it hep-ph/9703276}; \newref 
   Bl\"umlein J 1997 {\it Z.\ Phys.} C {\bf 74} 605; \newref
   Babu K S, Kolda C, March-Russell J and Wilczek F 1997 {\it Phys.\
   Lett.} {\bf B402} 367; \newref
   Hewett J L and Rizzo T G 1997 {\it Phys.\ Rev.} D {\bf 56} 5709; 
   \newref Kunszt Z and Stirling W J 1997 {\it Z.\ Phys.} C {\bf 75}
   453; 
   \newref Jadach S, P{\l}aczek W and Ward B F L 1997 {\it
     hep-ph/9705395} 

\bibitem{cao} Cao Z, He X-G and McKellar B 1997 {\it hep-ph/9707227} 

\bibitem{tev} Eno S (D0 Collaboration) 1997 {\it These proceedings};
   \newref Conway J (CDF Collaboration) 1997 {\it These proceedings}

\bibitem{D0} Abbott B \etal (D0 Collaboration) 1997 {\it
    Fermilab-Pub}-97/252-E

\bibitem{kraemer} Kr\"amer M, Plehn T, Spira M and Zerwas P M 1997 {\it
    Phys.\ Rev.\ Lett.} {\bf 79} 341

\bibitem{cdf} Abe F \etal (CDF Collaboration) 1997 {\it
    Fermilab-Pub}-97/280-E 

\bibitem{vector} Valls J A 1997 {\it Fermilab-Conf}-97/135-E, {\it XXXII 
    Rencontres de Moriond, QCD and Hadronic Interactions} (Les Arcs)  

\bibitem{lqst} Abe F \etal (CDF Collaboration) 1995 {\it Phys.\ Rev.\
    Lett.} {\bf 75} 1012; 1997 {\it Phys.\ Rev.\ Lett.} {\bf 78} 2906; 
    \newref Abbott B \etal (D0 Collaboration) 1997 {\it Paper 109
      submitted to the International Europhysics Conference on High
      Energy Physics} (Jerusalem)

\bibitem{h1lq} Aid S \etal (H1 Collaboration) 1996 {\it Phys.\ Lett.}
  {\bf B369} 173
  
\bibitem{zeuslq} Derrick M \etal (ZEUS Collaboration) 1993 {\it Phys.\
    Lett.} {\bf B306} 173; 1997 {\it Z.\ Phys.} C {\bf 73} 613

\bibitem{lqcontact} Haberl P, Schrempp F and Martyn H-U 1991 
  {\it Proceedings Physics at HERA}
  eds. Buchm\"uller W and Ingelman G (DESY, Hamburg)

\bibitem{h194} Ahmed T \etal (H1 Collaboration) 1994  {\it Z.\ Phys.}
  C {\bf 64} 545

\bibitem{leps} OPAL Collaboration 1997 {\it Paper LP138 submitted to the
    XVIII International Symposium on Lepton-Photon Interactions}
  (Hamburg)

\bibitem{lepp}
  Adeva B \etal (L3 Collaboration) 1991 {\it Phys.\ Lett.} {\bf B261}
  169; \newref
  Alexander G \etal (OPAL Collaboration) 1991 {\it Phys.\ Lett.} {\bf
    B263} 123; \newref 
  Abreu P \etal (DELPHI Collaboration) 1993 {\it Phys.\ Lett.} {\bf
    B316} 620

\bibitem{eeqq} Ackerstaff K \etal (OPAL Collaboration) 1997
  {\it CERN-PPE}/97-101

\bibitem{effl} Eichten E, Lane K and Peskin M 1983  
   {\it Phys.\ Rev.\ Lett.} {\bf 50} 811; \newref 
   R\"uckl R 1983 {\it Phys.\ Lett.} {\bf B129} 363;
             1984 {\it Nucl.\ Phys.} {\bf B234} 91

\bibitem{ks} Kunszt Z and Stirling W J 1997 {\it Z.\ Phys.} C {\bf 75}  
  453

\bibitem{agm} Altarelli G, Giudice G F and Mangano M L 1997 
  {\it hep-ph/9705287}

\bibitem{babu} Babu K S, Kolda C and March-Russell J 1997 
  {\it Phys.\ Lett.}\ {\bf B408} 261

\bibitem{hr} Hewett J L and Rizzo T G 1997 {\it hep-ph/9708419}

\bibitem{squarks} Choudhury D and Raychaudhuri S 1997 {\it Phys.\
    Lett.} {\bf B401} 54; \newref
  Altarelli G, Ellis J, Giudice G F, Lola S and Mangano M L 1997
  {\it hep-ph/9703276}; \newref  
  Dreiner H and Morawitz P 1997 {\it Nucl.\ Phys.}\ B {\bf 503} 55; 
  \newref
  Kalinowski J, R\"uckl R, Spiesberger H and Zerwas P M 1997 {\it Z.\
  Phys.} C {\bf 74} 595; \newref
  Kon T and Kobayashi T 1997 {\it hep-ph/9704221}; \newref
  Ellis J, Lola S and Sridhar K 1997 {\it Phys.\ Lett.}\ {\bf B408} 252; 
  \newref
  Kon T, Matsushita T and Kobayashi T 1997 {\it hep-ph/9707355}; \newref
  Carena M, Choudhury D , Raychaudhuri S and Wagner C E M 1997
  {\it hep-ph/9707458}

\bibitem{rp} Hewett J L 1992 {\it Summer Study
   on High Energy Physics -- Research Directions for the Decade
  (Snowmass 1990)}, ed.\ Berger E (World Scientific); \newref
  Kon T, Nakamura K and Kobayashi T 1990 {\it Z.\ Phys.} C {\bf 45}
  567; 1990 {\it Acta Phys.\ Polon.} B {\bf 21} 315; \newref
  Butterworth J and Dreiner H 1993 {\it Nucl.\ Phys.} B {\bf 397} 3;
  \newref Dreiner H, Perez E and Sirois Y 1996 {\it Future Physics at
    HERA} eds.\ Ingelman G, De Roeck A and Klanner R (DESY, Hamburg) 

\bibitem{dreiner2} Dreiner H 1997 {\it hep-ph/9707435}  

\bibitem{kon} Kon T and Kobayashi T 1997 {\it hep-ph/9704221}

\bibitem{pp2} R\"uckl R and Zerwas P M 1987 {\it Physics at Future
  Accelerators} ed.\ Mulvey J H (La Thuile, Geneva) {\it CERN report}
  87-07; \newref 
  Bhattacharyya G, Choudhury D and Sridhar K 1995 {\it
  Phys.\ Lett.} {\bf B349} 118

\end{thebibliography}
\end{document}